\documentclass[twocolumn]{aa}
\usepackage{natbib}
\usepackage{units}
\usepackage{graphicx}
\usepackage{subfigure}
\usepackage{mathtools}
\usepackage{twoopt}
\usepackage{url}
\usepackage{listings}

\bibpunct{(}{)}{;}{a}{}{,} 
\makeatletter
\newcommandtwoopt{\citeads}[3][][]{\href{http://adsabs.harvard.edu/abs/#3}%
{\def\hyper@linkstart##1##2{}%
\let\hyper@linkend\@empty\citealp[#1][#2]{#3}}}
\newcommandtwoopt{\citepads}[3][][]{\href{http://adsabs.harvard.edu/abs/#3}%
{\def\hyper@linkstart##1##2{}%
\let\hyper@linkend\@empty\citep[#1][#2]{#3}}}
\newcommandtwoopt{\citetads}[3][][]{\href{http://adsabs.harvard.edu/abs/#3}%
{\def\hyper@linkstart##1##2{}%
\let\hyper@linkend\@empty\citet[#1][#2]{#3}}}
\newcommandtwoopt{\citeyearads}[3][][]%
{\href{http://adsabs.harvard.edu/abs/#3}
{\def\hyper@linkstart##1##2{}%
\let\hyper@linkend\@empty\citeyear[#1][#2]{#3}}}
\makeatother

\newcommand{\lsi}{LS~I~$+61^{\circ}303$}

\setcitestyle{notesep={; },round,aysep={},yysep={;}}

\begin{document}

\title{
  Discovery of a periodical apoastron GeV peak in 
   \lsi{}
}

\author{F.~Jaron and M.~Massi}

\institute{
  Max-Planck-Institut\ f\"ur\ Radioastronomie,
  Auf dem H\"ugel 69,
  D-53121 Bonn, Germany \\ 
  \email{fjaron, mmassi@mpifr-bonn.mpg.de}
}

\abstract
{}
{ 
  The aim of this paper is to analyse the previously discovered
  discontinuity of the periodicity of the GeV $\gamma$-ray emission of
  the radio-loud X-ray binary\ \lsi{} and to determine its physical
  origin.
}
{ 
  We used wavelet analysis to explore the temporal development of
  periodic signals. The wavelet analysis was first applied to the whole
  data set of available \textit{Fermi}-LAT data and then to the two
  subsets of orbital phase intervals $\Phi = 0.0 - 0.5$ and $\Phi =
  0.5 - 1.0$. We also performed a Lomb-Scargle timing analysis. We
  investigated the similarities between GeV $\gamma$-ray emission and
  radio emission by comparing the folded curves of the
  \textit{Fermi}-LAT data and the Green Bank Interferometer radio
  data.
}
{ 
  During the epochs when the timing analysis fails to determine the
  orbital periodicity, the periodicity is present in the two orbital
  phase intervals\ $\Phi = 0.0 - 0.5$ and $\Phi = 0.5 - 1.0$. That is,
  there are two periodical signals, one towards periastron (i.e.,
  $\Phi = 0.0 - 0.5$) and another one towards apoastron ($\Phi = 0.5 -
  1.0$). The apoastron peak seems to be affected by the same orbital
  shift as the radio outbursts and, in addition, reveals the same two
  periods\ $P_1$ and\ $P_2$ that are present in the radio data.
}
{ 
  The $\gamma$-ray emission of the apoastron peak normally just
  broadens the emission of the peak around periastron. Only when it
  appears at $\Phi \approx 0.8 - 1.0$, because of the orbital shift,
  it is enough detached from the first peak to become recognisable as
  a second orbital peak, which is the reason why the timing analysis
  fails. Two $\gamma$-ray peaks along the orbit are predicted by the
  two-peak  accretion model for an eccentric orbit, that was proposed
  by several authors for \lsi{}.
}


\titlerunning{  
  Discovery of a periodical apoastron GeV peak in \lsi{}
}

\maketitle

\section{Introduction}

The system \lsi{} with an orbital period \ $P_1= \unit[26.4960 \pm
  0.0028]{days}$ \citep{Gregory2002} consists of a compact object and
a massive star with an optical spectrum typical for a rapidly rotating
B0 V star \citep{Casares2005, Grundstrom2007}.

In 2009 the first detection of orbital periodicity in high-energy
gamma rays (20 MeV–100 GeV) was reported by using the Large Area
Telescope (LAT) from the \textit{Fermi} Gamma-Ray Space Telescope
spacecraft \citep{Abdo2009}. Longer monitoring has shown
\citep{Hadasch2012, Ackermann2013} that indeed the system shows a
clear periodical outburst towards periastron \citep[$\Phi_{\rm
    periastron} = 0.23$,][]{Casares2005} at some epochs, but that this
periodicity is not always present. This is different from the
behaviour of the system in the radio band where not only periodical
outbursts occur at each orbit, even if modulated with a long-term
period \citep[$P_{\rm long}$ = 1667 $\pm$ 8~d,][]{Gregory2002}, but
they also occur towards apoastron and not towards periastron as in the
GeV energy band \citep[e.g., see Fig.~2\,c in][]{Massi2009}.

Along with this different behaviour between high and low energy there
is a puzzling overlap. \citet{Ackermann2013} noticed  that 
GeV data also show the long-term periodical variation  affecting
the radio data, but only at a specific orbital phase interval, $\Phi =
0.5 - 1.0$, that is around apoastron.

The aim of this paper is to investigate the discontinuity in the
periodicity of the GeV $\gamma$-ray emission at periastron, the
relationship of its disappearance with the variation of the emission
in other parts of the orbit, and finally the possible relationship
between GeV and radio emission. Section~2 describes our data
analysis. In Sect. 3 we present our results and in Sect.\,4 our
conclusions.

 
\section{Data analysis}

In this section we present the data reduction of \textit{Fermi}-LAT
data performed with three packages: the \textit{Fermi} science tools
package (version
v9r33p0)\footnote{\url{http://fermi.gsfc.nasa.gov/ssc/data/analysis/software/}},
the wavelet
analysis\footnote{\url{http://atoc.colorado.edu/research/wavelets/}},
and the software
Starlink\footnote{\url{http://www.starlink.rl.ac.uk/}}.

The $\gamma$-ray data used in this analysis span the time period
MJD~54683 (August 05, 2008) to MJD~56838 (June 30, 2014). We used the
script like\_lc.pl by
Robin~Corbet.\footnote{\url{http://fermi.gsfc.nasa.gov/ssc/data/analysis/user/}}
Only source-event-class photons were selected for the
analysis. Photons with a zenith angle greater than 100$^{\circ}$ were
excluded to reduce contamination from the Earth's limb. For the
diffuse emission we used the model gll\_iem\_v05\_rev1.fit and the
template iso\_source\_v05\_rev1.txt. We used the instrument response
function (IRF) P7REP/background\_rev1, and the model file was
generated from the 2FGL catalogue \citep{Nolan2012}, all sources
within $10^{\circ}$ of \lsi{} were included in the model. \lsi{} was
fitted with a log-parabola spectral shape and with all parameters left
free for the fit, performing an unbinned maximum likelihood
analysis. The other sources were fixed to their catalogue values. We
produced light curves with a time bin size of one day and of five
days.
For all light curves we used an energy range of 100~MeV to 300~GeV.

The data were folded with the orbital phase defined as
\begin{equation}
  \label{eq:Phi}
  \Phi = \frac{t - t_0}{P_1} - {\rm int}\left(\frac{t -
    t_0}{P_1}\right),
\end{equation}
where $t_0 = \unit[43366.275]{MJD}$ and $P_1= \unit[26.4960 \pm
  0.0028]{d}$  is the orbital period of the binary system
\citep{Gregory2002}.

\begin{figure}
  \includegraphics{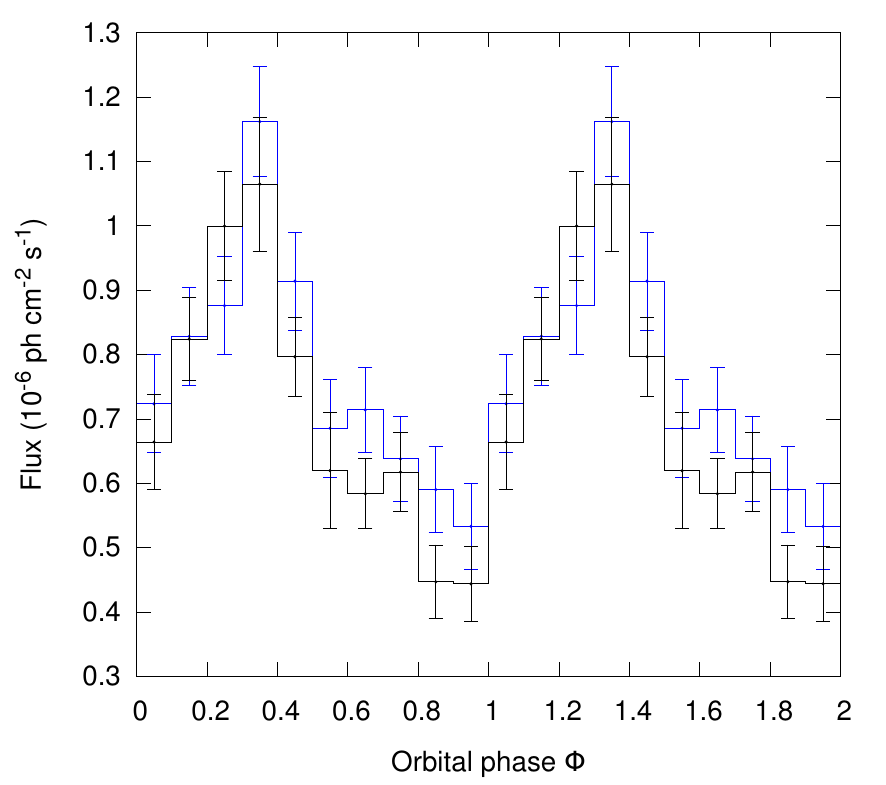}
  \caption{
    Folded light curve (blue) of Fig.~8 by \citet{Hadasch2012}. This
    covers the first 8 months of Fermi-LAT observations. As comparison
    we give our folded data (black) from the same time interval. This
    time interval corresponds to $\Theta = 6.79 - 6.92$.
  }
  \label{fig:First8months} 
\end{figure}

\begin{figure*}[Ht]
  \subfigure[]{
    \hbox{\hspace{-.66em}
      \includegraphics{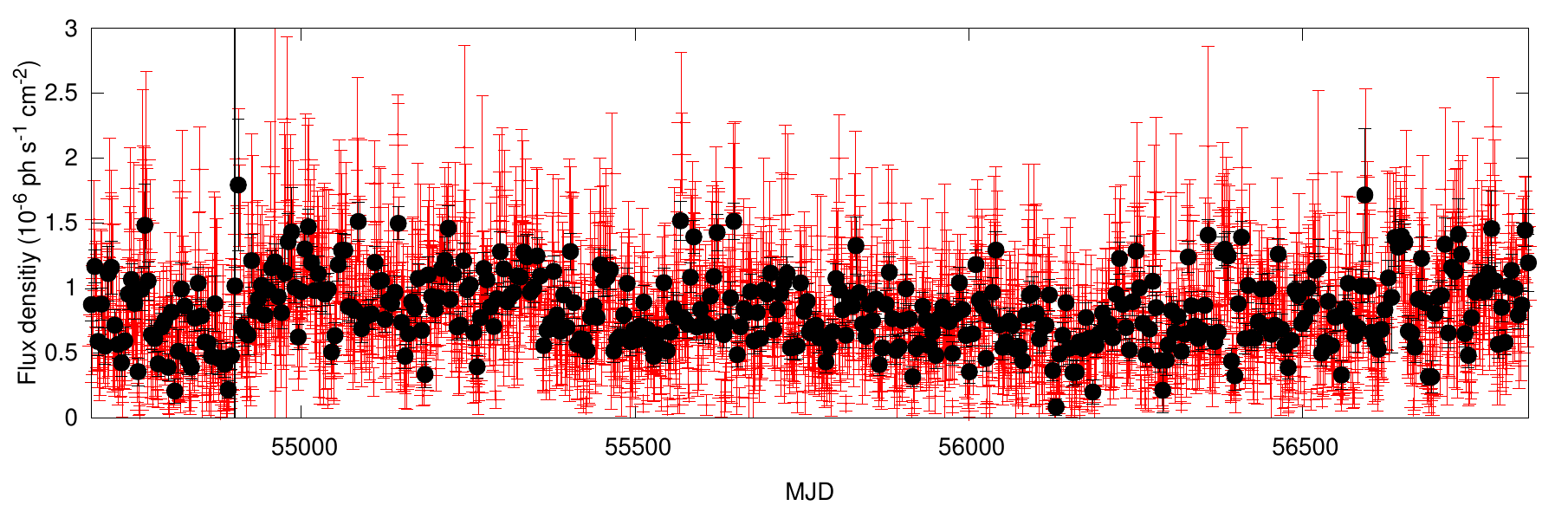}
    }
  }
  \subfigure[]{
    \includegraphics{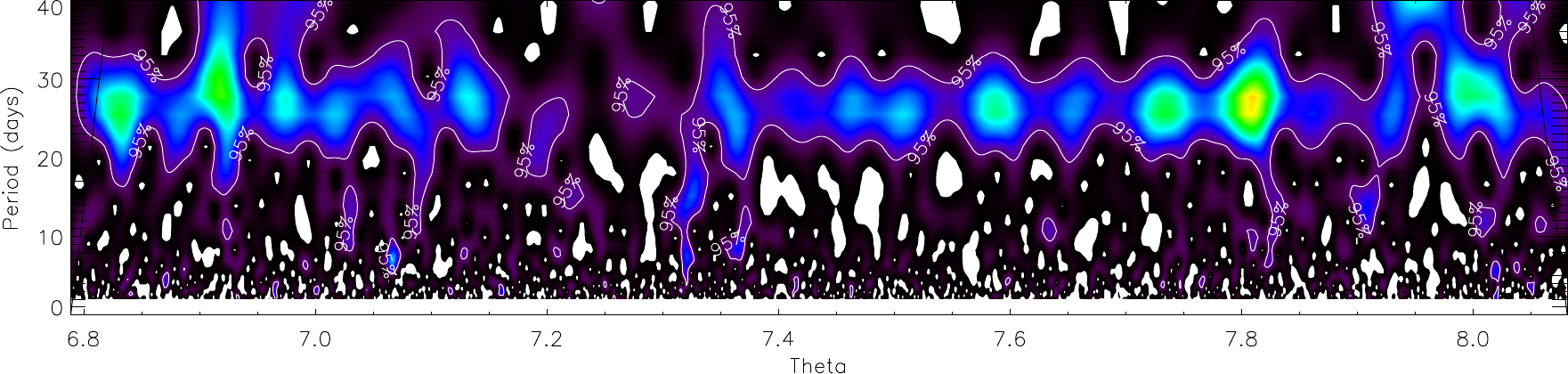}
  }
  \subfigure[]{
    \includegraphics{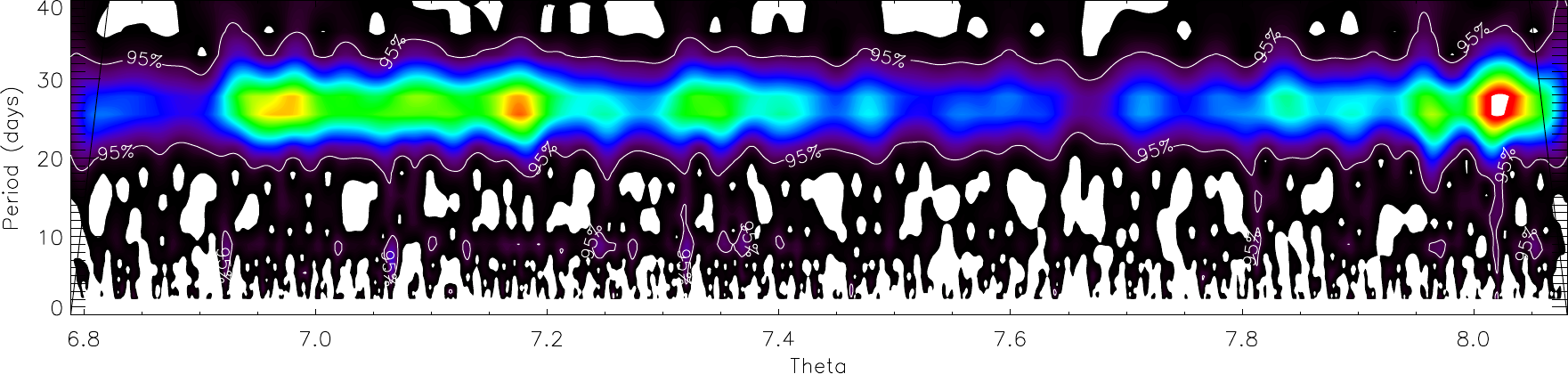}
  }
  \subfigure[]{
    \includegraphics{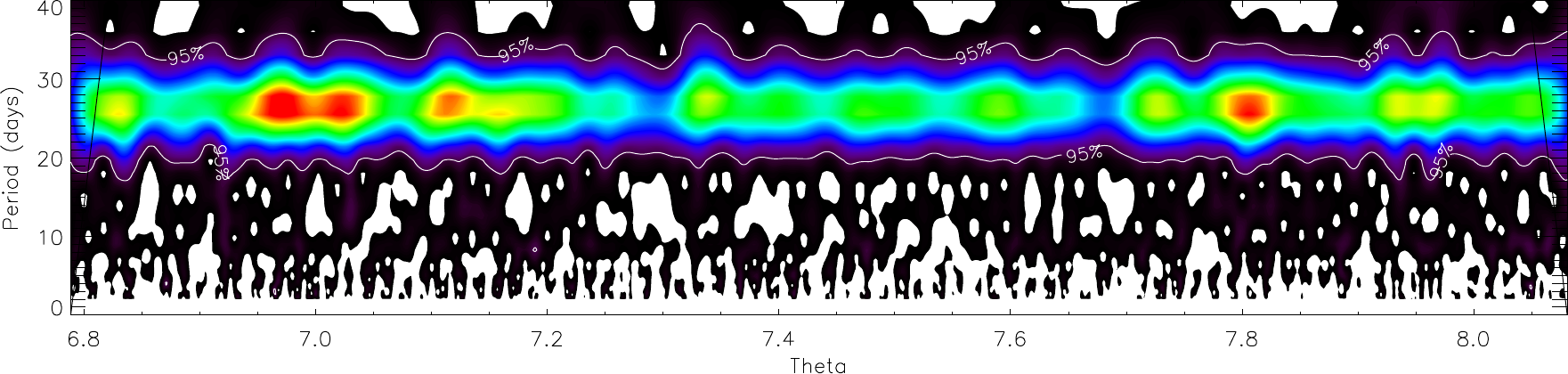}
  }
  \subfigure{
    \includegraphics{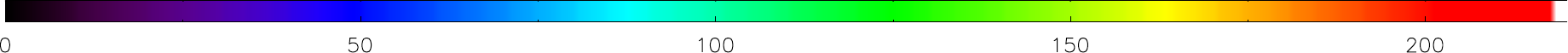}
  }
  \caption{
    Wavelet analysis of \textit{Fermi}-LAT data. The strength of
    periodicity is colour coded as indicated in the bottom bar.
    (a) \textit{Fermi}-LAT data with a time bin of 5~d (black)
    are overplotted on \textit{Fermi}-LAT data with a time bin of 1 d (red).
The black line marks the
 point of flux change reported by  \citet{Hadasch2012} which is also
 visible in our light curve.
    (b) Wavelet analysis for the whole 
data set, that is for the whole
orbital interval $0.0-1.0$ (b--d use a
    time bin of one day).
    (c) Wavelet analysis for half  the data set, that is for
the orbital interval\ $\Phi = 0.5 -1.0$,
    i.e., around apoastron.
    (d) Wavelet for half  the data set, that is for the orbital interval\ $\Phi = 0.0 -0.5$, i.e.,
    around periastron.
  }
  \label{fig:FermiLATWL}
\end{figure*}

We investigated the temporal evolution of the orbital periodicity by
means of a wavelet analysis with Morlet function
\citep{Torrence1998}. The wavelet analysis decomposes the
one-dimensional time series into a two-dimensional time-frequency
space and displays the power spectrum in a two-dimensional colour-plot
that shows how the Fourier periods vary in time \citep{Torrence1998}.
While the wavelet analysis was applied to the $\gamma$-ray data vs
time, for a straightforward comparison with radio data, we express the
$x$-axis as
\begin{equation}
  \label{eq:Theta}
  \Theta = \frac{t - t_0}{P_{\rm long}}.
\end{equation}
This allows a comparison with non-simultaneous radio data because the
radio data are periodical in $\Theta$. We will therefore compare
gamma-ray data to radio data having the same fractional part of
$\Theta$. For the Lomb-Scargle timing analysis \citep{Lomb1976,
  Scargle1982},
we used the program PERIOD, which is part of the UK software
Starlink.
The version we used was 5.0-2 for UNIX. The wavelet analysis assumes
regularly sampled data. We therefore set the data for the wavelet
analysis to zero for missing flux. For the Lomb-Scargle analysis this was
not necessary. As discussed in Sect. 3, the Lomb-Scargle
analysis confirms and accurately determins the periodicities found
with the wavelet analysis. In the Lomb-Scargle and wavelet analysis,
significance levels for the spectra were determined with the
Fisher randomisation, as outlined in \citet{LinnellNemec1985}, and
with Monte Carlo simulations, as in \citet{Torrence1998}.
The fundamental assumption is: if there is no periodic signal in the
time series data, then the measured values are independent of their
observation times and are likely to have occurred on any other
order. One thousand randomized time-series were formed and the
periodograms calculated. The proportion of permutations that give a
peak power higher than that of the original time series would then
provide an estimate of $p$, the probability that for a given frequency
window there is no periodic component present in the data with this
period. A derived period is defined as significant for $p < 0.01$, and
a marginally significant period for $0.01 < p < 0.10$
\citep{LinnellNemec1985}.


\section{Results}

In this section we discuss our wavelet and Lomb-Scargle results for
the gamma-ray data and compare them with previously published results
\citep{Hadasch2012, Ackermann2013}. Then we compare gamma-ray data
with radio data.

\subsection{Wavelet and Lomb-Scargle analysis}

As a general check of our data reduction we verified the consistency
of the folded curve of our data with the folded curve of
\citet{Hadasch2012}. For this purpose we selected the same time
interval as in \citet{Hadasch2012}, that is 54683--54900~MJD ($\Theta
= 6.79 - 6.92$), covering the first eight months of \textit{Fermi}-LAT
observations. Figure~1 presents the result by \citet{Hadasch2012} and
our results. We see how these independently calibrated data sets,
which used different versions of the software, of the instrumental
response funtion, etc., produce the same result: There is a main peak
at orbital phase $\Phi \approx 0.35$.

We compare the light curve and wavelet results with previous
results. The whole interval of the \textit{Fermi}-LAT data used in our
analysis is presented in Fig.~\ref{fig:FermiLATWL}\,a.
The point of flux change reported by  \citet{Hadasch2012}  is also
 visible in our light curve.
Figure~\ref{fig:FermiLATWL}\,b shows the wavelet analysis results for
the whole data set. The orbital periodicity shows a minimum around
$\Theta \approx 7.25$. This agrees with the periodograms of Fig.~7 in
\citet{Hadasch2012}, where the orbital periodicity is already almost
absent for MJD~55044-55225 ($\Theta = 7.00 - 7.11$) and is completely
absent at MJD~55405-55586 ($\Theta = 7.22 - 7.33$). The panels of
Fig.~4 in \citet{Ackermann2013} show the 
decline in the orbital flux modulation in the
interval\ MJD~55191--55698 ($\Theta = 7.1 - 7.5$). Our Figs.~1, 2\,a
and 2\,b therefore confirm the orbital modulation that peaks around
periastron, the point of flux change at $\Theta \approx 6.95$, and the lack of
orbital flux modulation  at $\Theta \approx 7.25$ previously
discovered and discussed in \citet{Hadasch2012} and
\citet{Ackermann2013}.

Now we examine the new results obtained when the wavelet analysis is
performed separately for emission around periastron, $\Phi = 0.0 -
0.5$, and emission around apoastron, $\Phi = 0.5 - 1.0$. By comparing
Fig.~\ref{fig:FermiLATWL}\,d with Fig.~\ref{fig:FermiLATWL}\,c it is
clear that the orbital periodicity is not a characteristic of
periastron emission alone, it is also present at apoastron, at least
at some $\Theta$s. At $\Theta \approx 7.2$ where the wavelet analysis
for the whole data set (Fig.~\ref{fig:FermiLATWL}\,b) shows the
minimum power for the orbital periodicity, a maximum  is
present in Fig.~\ref{fig:FermiLATWL}\,c. Orbital flux modulation is
present from $\Theta \approx$ 6.95 to $\Theta \approx$ 7.40, and at
$\Theta \approx 7.95$ the periodicity is strong again, indicating a
repetition consistent with the long-term period\ $P_{\rm long}$. This
confirms the discovery of \citet{Ackermann2013} that in the orbital
phase interval\ $\Phi = 0.5 - 1.0$ the $\gamma$-ray flux shows
periodical variation with a period equal to the long-term modulation
affecting the radio outburst. In addition, our
wavelet analysis shows that the emission at apoastron, affected by the
long-term modulation, is orbitally modulated.

\begin{figure*}
  \subfigure[]{
    \includegraphics{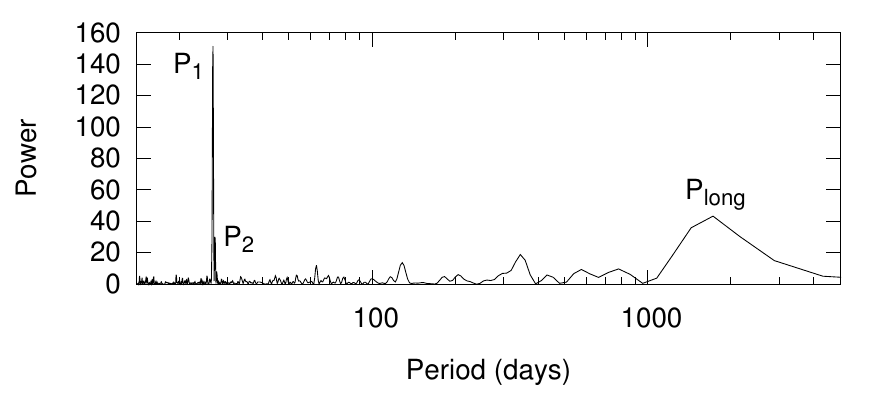}
  }
  \subfigure[]{
    \includegraphics{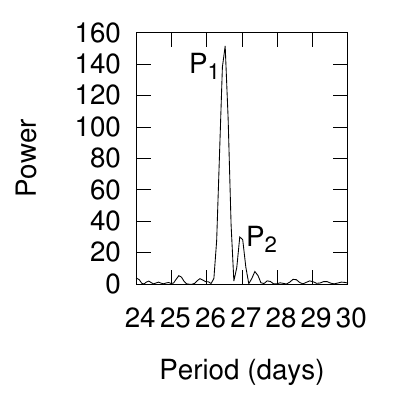}
  }
 \subfigure[]{
    \includegraphics{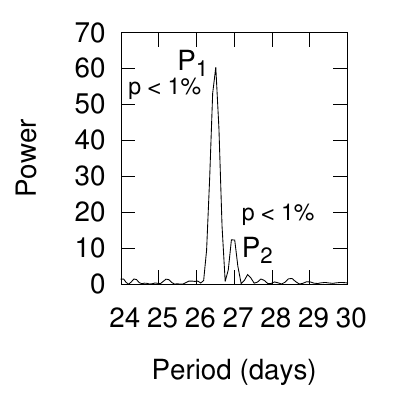}
  }\\
  \subfigure[]{
    \includegraphics{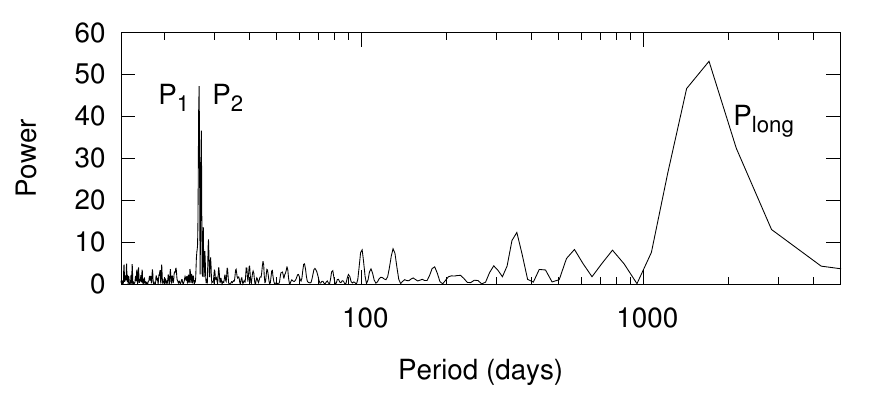}
  }
  \subfigure[]{
    \includegraphics{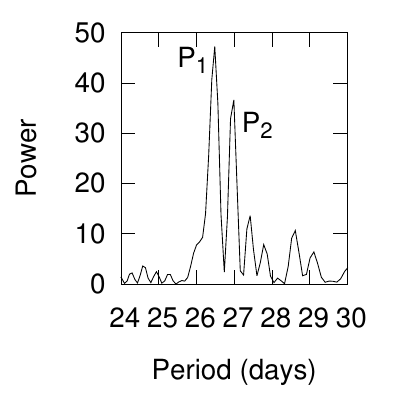}
  }
  \subfigure[]{
    \includegraphics{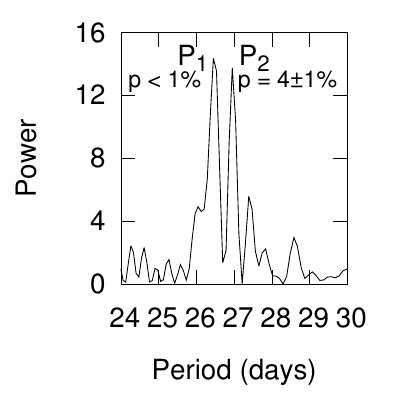}
  }\\
  \subfigure[]{
    \includegraphics{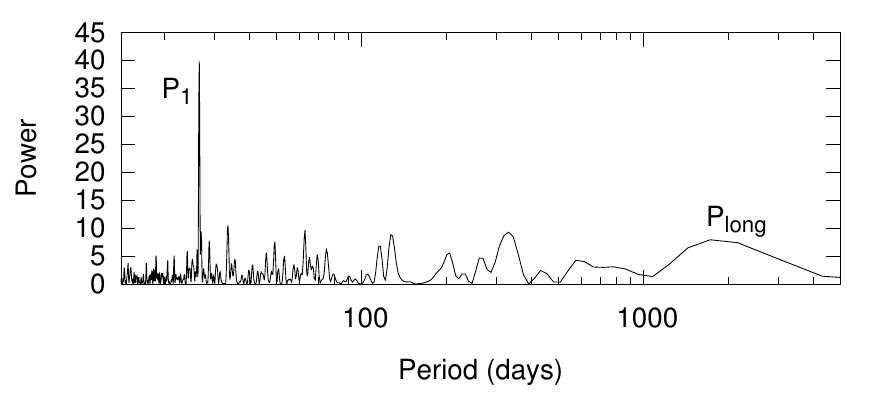}\\
  }
  \subfigure[]{
    \includegraphics{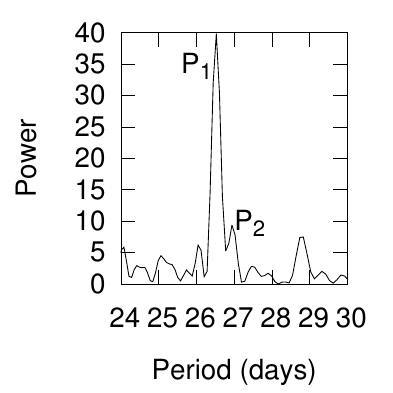}
  }
  \subfigure[]{
    \includegraphics{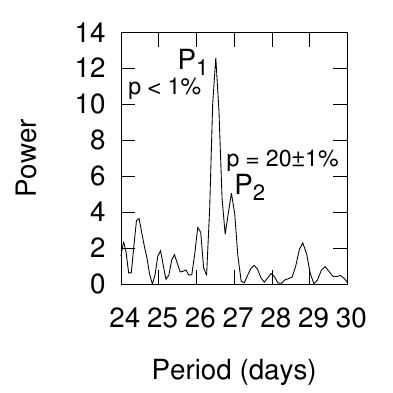}
  }\\
  \caption{
    Lomb-Scargle periodogram of the \textit{Fermi}-LAT (with a time
    bin of one day).
    (a) Full data set: data in the orbital phase $\Phi = 0.0 - 1.0$.
    (b) Zoom of Fig.~3\,a.
    (c) Same as 3\,b for data with a time bin of 5 d.
    The false alarm probability ($p$) resulting from randomisation tests 
    is indicated. A period is defined as significant for $p < 1\%$,
    and as marginally significant for $1\% < p < 10\%$ (Sect.~2).
    (d) Half the data set: data in the orbital phase $\Phi = 0.5 - 1.0$.
    The periods\ $P_2$ and\ $P_{\rm long}$ here present are typical
    periodicities in radio data \citep{Massi2013}. 
    (e) Zoom of Fig.~3\,d.
    (f) Same as 3\,e for data with a time bin of 5~d.
    (g) Half the data set: data in the orbital phase $\Phi = 0.0 - 0.5$.
    (h) Zoom of Fig.~3\,g.
    (i) Same as 3\,h for data with a time bin of 5~d.
  }
  \label{fig:FermiLATSC}
\end{figure*}

We checked the presence of the orbital and long-term periodicities
that were indicated by the wavelet analysis with the Lomb-Scargle
timing analysis of the data from $\Phi = 0.5-1.0$. First of all, the
periodogram of the subset $\Phi = 0.5-1.0$, shown in
Fig.~\ref{fig:FermiLATSC}\,d, presents a periodicitiy at $1705 \pm
355$~days, confirming the result of \citet{Ackermann2013} of a
long-term modulation of emission around apoastron. Moreover, the
periodogram shows a feature at the orbital period $P_1$, confirming
the results of the wavelet analysis, and this feature seems to have a
complex profile. The zoomed version is presented in
Fig.~\ref{fig:FermiLATSC}\,e. Figure~\ref{fig:FermiLATSC}\,f shows the
periodogram for data integrated over five days. There is one peak at
$P_1 = \unit[26.48 \pm 0.08]{d}$ and another one at $P_2 = \unit[26.99
  \pm 0.08]{d}$. When we consider the whole data set, which is the
longest time interval of \textit{Fermi}-LAT data analysed so far,
$P_{\rm long}$ and $P_2$ are significant in the randomisations tests
(i.e., the probability $p$ that there is no periodic component present
in the data with these periods is for both periods $p < 0.01$), even
if they are rather weak features  in the periodograms of
Fig.~\ref{fig:FermiLATSC}\,a, b, and c. For the emission around
periastron, Figures~\ref{fig:FermiLATSC}\,g, h, and i show the
Lomb-Scargle periodograms of $\Phi = 0.0 - 0.5$. Here the only
outstanding feature is $P_1$.
 
\subsection{Orbital shift}

\begin{figure*}
  \subfigure[]{
    \includegraphics{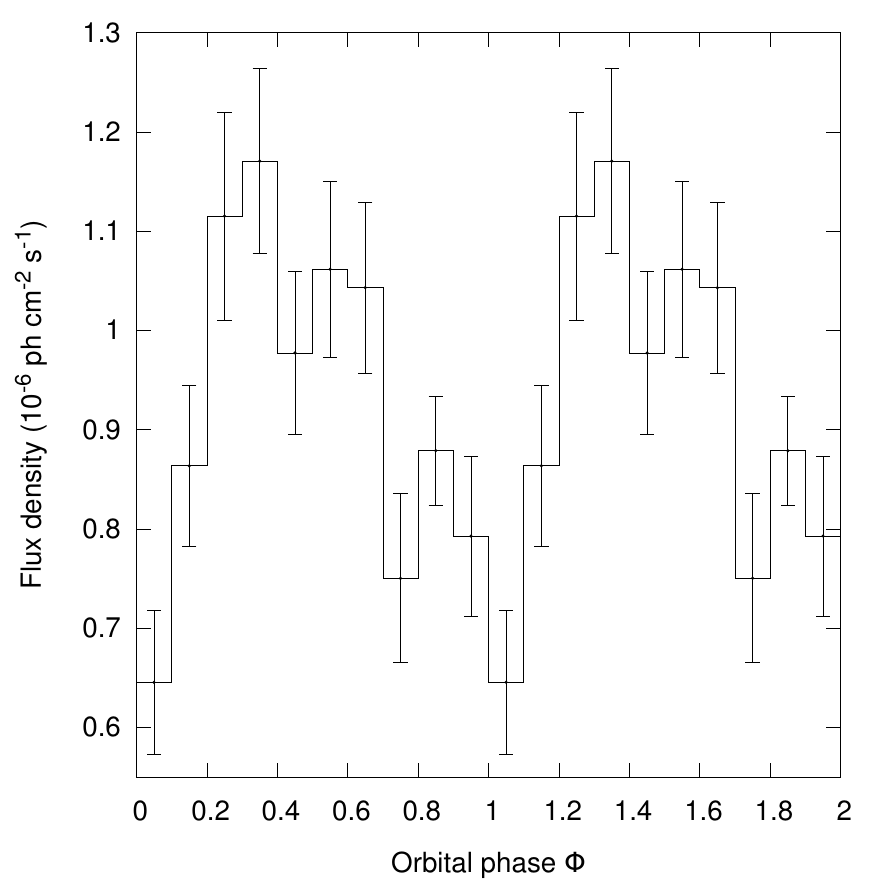}
  }
  \subfigure[]{
    \includegraphics{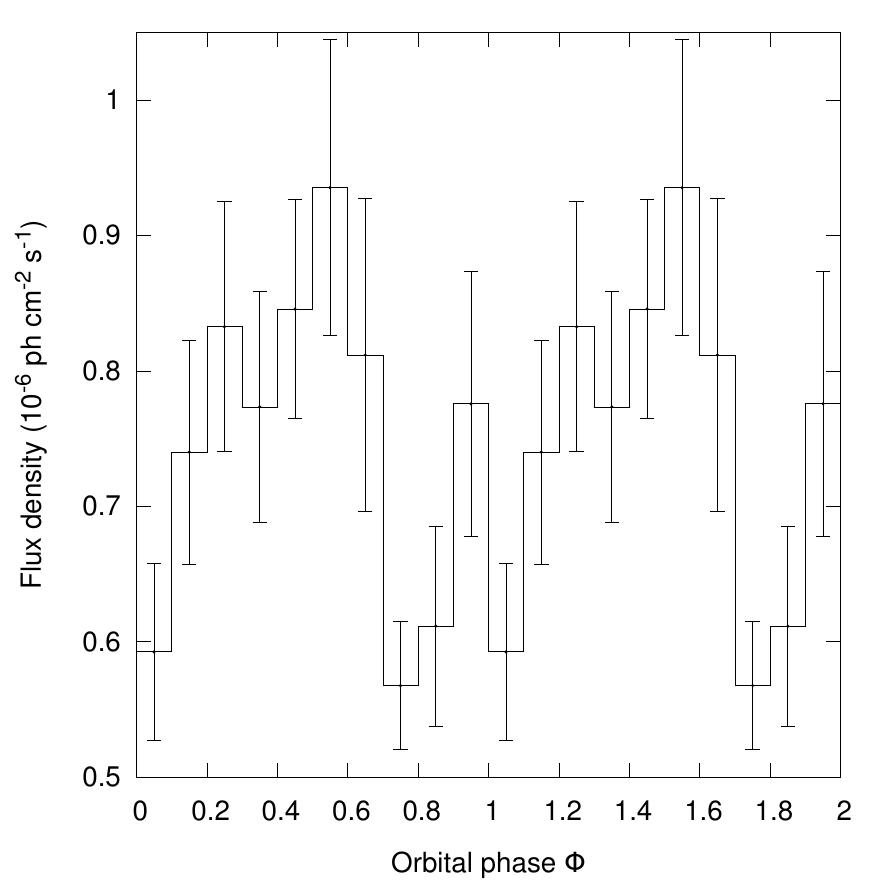}
  }
  \subfigure[]{
    \includegraphics{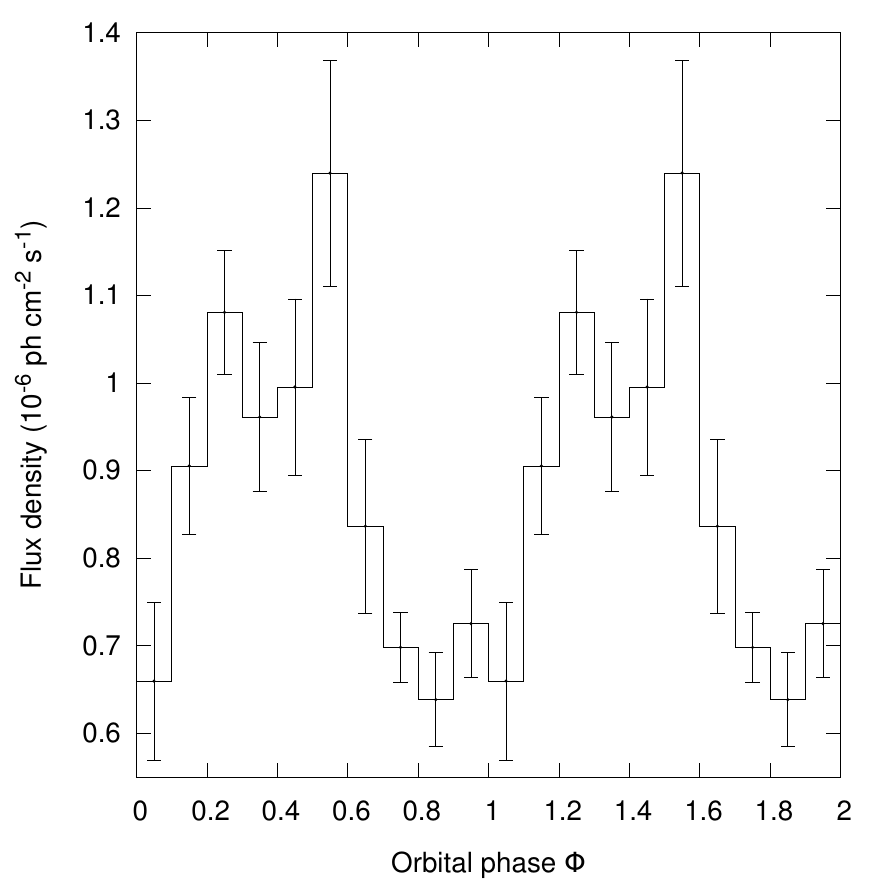}
  }
   \subfigure[]{
  \includegraphics{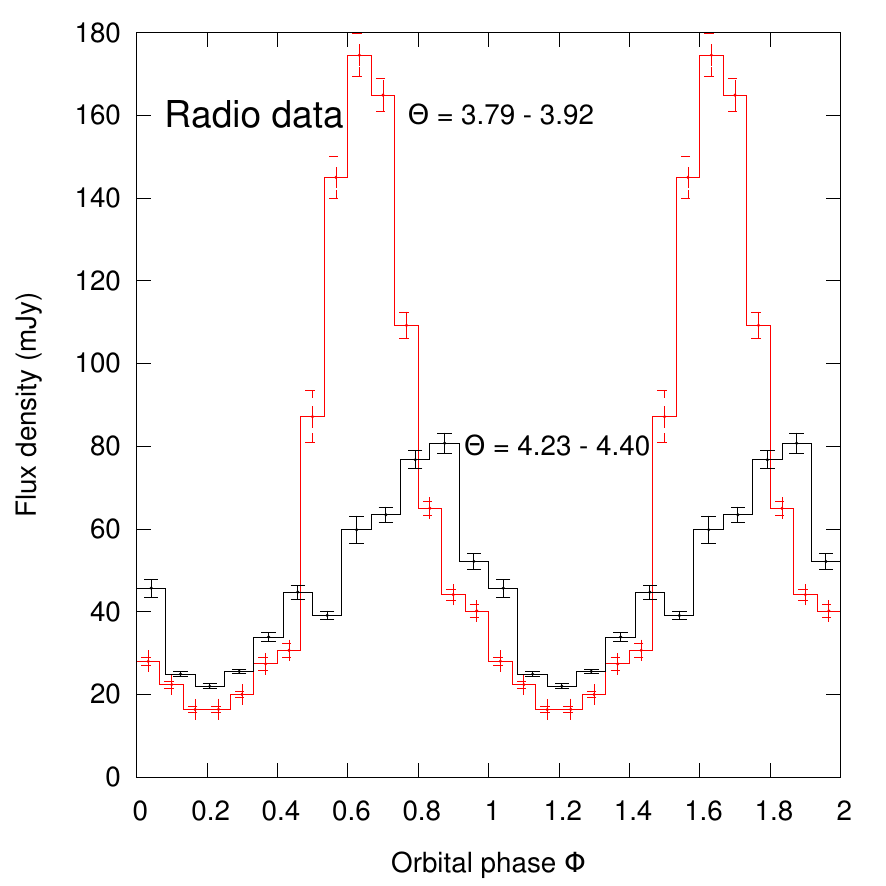}
   }
  \caption{
    Apoastron peak. \textit{Fermi}-LAT gamma-ray data folded with
    the orbital periodicity\ $P_1 = \unit[26.4960]{d}$ for three
    consecutive $\Theta$-intervals.
    (a) $\Theta = 7.12 - 7.22$ (MJD~55235--55402);
    (b) $\Theta = 7.22-7.32$ (MJD~55402--55569);
    (c) $\Theta = 7.32-7.42$ (MJD~55569--55740).
    (d) 
    GBI radio data at 8~GHz folded with orbital period $P_1 =
    \unit[26.4960]{d}$. The two $\Theta$-intervals are those
    considered in Fig.~\ref{fig:First8months} and in this figure
    for \textit{Fermi}-LAT data. The orbital phase of the peak of the
    two radio curves agrees well with those of the second peak in
    GeV curves.
  }
  \label{fig:2ndpeak}
\end{figure*}

Why does the timing analysis reveal a lack of orbital modulation
around $\Theta \approx 7.2$ with a peak at periastron passage when the
periodicity is still present, as shown in Fig.~\ref{fig:FermiLATWL}\,d
for data at $\Phi = 0.0 - 0.5$? And why does the orbital modulation of
GeV emission towards apoastron get stronger in that $\Theta$-interval
(Fig.~\ref{fig:FermiLATWL}\,c)? Clearly, the second question is the
answer to the first. Two peaks along the orbit disturb the timing
analysis. The curves shown in Fig.~\ref{fig:2ndpeak}\,a, b, and c
refer to the three consecutive $\Theta$-intervals around the minimum
of Fig.~\ref{fig:FermiLATWL}\,b, that is around the peak of
Fig.~\ref{fig:FermiLATWL}\,c: $\Theta = 7.12 - 7.22$
(MJD~55235--55402), $\Theta = 7.22 - 7.32$ (MJD~55402--55569), and
$\Theta = 7.32 - 7.42$ (MJD~55569--55743). In addition to the peak at
periastron Fig.~\ref{fig:2ndpeak}\,a and Fig.~\ref{fig:2ndpeak}\,b
show a second peak in the interval $\Phi = 0.8 - 1.0$.

The real question therefore is why periodical emission towards
apoastron is detected only at $\Theta \approx 7.2$. The important
information from the Lomb-Scargle analysis is that there are two
periods, $P_1$ and $P_2$, as is for radio data \citep{Massi2013}. We
therefore examined the trend of radio data in that particular
$\Theta$-interval. Figure~4\,d shows the GBI data  
at 8~GHz at $\Theta = 4.23 - 4.40$. The curve shows indeed a peak at
$\Phi \approx 0.9$ consistent with the second peak in the GeV data in
the interval  $\Phi = 0.8 - 1.0$ 
(Fig.~\ref{fig:2ndpeak} a-b). Figure~4~d also shows data at $\Theta =
3.79-3.92$, as the GeV data of our Fig.~\ref{fig:First8months}. In
this case, the radio data show a peak at $\Phi \approx 0.65$
consistent with the bump of emission at $\Phi \approx 0.65$ in the
gamma-ray data of Fig.~1. This is the well-known phenomenon of the
orbital shift of the radio outburst in \lsi{}: The largest outbursts
occur at orbital phase 0.6, afterwards, with the long-term
periodicity, the orbital phase of the peak of the outburst changes, as
analysed by \citet{Paredes1990b} in terms of orbital phase shift, by
\citet{Gregory1999} in terms of timing residuals, and reproduced
recently by the precessing jet model in \citet[their
  Fig. 4\,b]{Massi2014}.


The second $\gamma$-ray peak, clearly associated with the radio outburst in
terms of timing analysis, therefore also follows the same orbital
shift. 
The second peak  is therefore only detected
at $\Theta \approx 7.2$, although the periodicity is indeed always
present, because in that $\Theta$-interval the peak is detached enough
from the first $\gamma$-ray peak to be discernable in the timing
analysis.

\section{Discussion and conclusion}

The first eight months of the \textit{Fermi}-LAT observations ($\Theta
= 6.79 - 6.93$) analysed by \citet{Hadasch2012} indicate a peak of GeV
$\gamma$-ray emission close to periastron.
This clear orbital modulation of the GeV gamma-ray emission is lost at
some time interval \citep{Hadasch2012, Ackermann2013}. The aim of this
paper was to investigate the origin of this disappearance. Our results
are the following:
\begin{enumerate}
\item{
  In the interval $\Theta \approx 7.2$ (Fig.~\ref{fig:FermiLATWL}\,b)
  where the timing analysis fails to find the orbital periodicity in
  GeV $\gamma$-ray emission, there are two periodical signals, one in
  the orbital phase interval\ $\Phi = 0.0 - 0.5$
  (Fig.~\ref{fig:FermiLATWL}\,d), that is towards periastron, the
  other in the interval\ $\Phi = 0.5 - 1.0$
  (Fig.~\ref{fig:FermiLATWL}\,c), that is towards apoastron.
  This result of two GeV peaks along the orbit corroborates the
  two-peak accretion model for \lsi{}. The hypothesis that a compact
  object that accretes material along an eccentric orbit undergoes two
  accretion peaks along the orbit was suggested and developed by
  several authors for the system \lsi{} \citep{Taylor1992, Marti1995,
    Bosch-Ramon2006, Romero2007}. The first accretion peak is
  predicted to occur close to the Be star and to give rise to  a
  major high-energy outburst. The second accretion peak is predicted
  to occur much farther away from the Be star, where the radio
  outburst occurs, and a minor high-energy outburst is predicted there
  \citep{Bosch-Ramon2006}. The predicted periastron event corresponds
  well to the observed GeV peak towards periastron, the second
  predicted  high-energy outburst, corresponds well to the here
  discussed apoastron peak.
}
\item{
  The Lomb-Scargle analysis of emission around apoastron
  (Fig.~\ref{fig:FermiLATSC}\,d, e, f) revealed the same three
  periodicities $P_{1_{\gamma}} = \unit[26.48 \pm 0.08]{d}$,
  $P_{2_{\gamma}} = \unit[26.99 \pm 0.08]{d}$, and $P_{{\rm
      long}_{\gamma}} = \unit[1705 \pm 335]{d}$ that affect the radio
  emission: $P_{1_{\rm radio}} = \unit[26.4960 \pm 0.0028]{d}$,
  $P_{{\rm long}_{\rm radio}}= \unit[1667 \pm 8]{d}$
  \citep{Gregory2002}, and $P_{2_{\rm radio}}= \unit[26.92 \pm
    0.07]{d}$ \citep{Massi2013}.

  This second result confirms the previous result of $P_1$ in GeV
  emission around apoastron, and in addition
  reveals $P_2$, only recently discovered in the radio emission by the
  timing analysis of 6.7~years of Green Bank Radio Interferometer
  (GBI) observations at the two frequencies of 2.2\,GHz and 8.3\,GHz
  \citep{Massi2013}. The radio data base of GBI and the data base of
  \textit{Fermi}-LAT cover two quite different time intervals
  separated by 8~yr (the last scan in the GBI database is June 2000,
  whereas Fermi-Lat monitoring begins in August 2008). Moreover, the
  two monitorings have a quite different sampling rate (GBI with up to
  eight observations per day and large gaps, \textit{Fermi}-LAT covers
  the whole sky over three hours, and we integrated over one
  day). Nevertheless, the timing analysis gives the same three
  periodicities \citep[compare our Figs.~3\,d, e, f with Figs.~1 and
    2\,b in][]{Massi2013}.

  Figures 3\,g, h, and i reveal a quite different characteristic for
  emission around periastron: the Lomb-Scargle analysis results in
  only one outstanding feature at $P_1 = 26.52 \pm 0.08$ d. The
  connection of  $P_{\rm long}$ with  $P_2$ is evident from Figs.~3\,g
  and h: the lack of $P_2$ (or reduction at noise level) is associated
  with a lack of $P_{\rm long}$ (reduced at noise level). Simulations
  \citep{Massi2013} with $P_1$ and the long-term modulation cannot
  reproduce the observed periodogram (i.e., $P_2$), whereas
  simulations with $P_1$ and $P_2$ directly produce the long-term
  modulation as their beating frequency $\nu_1-\nu_2$.

  While $P_1$ is related to the periodical accretion peak towards
  apoastron described above, the period $P_2$ is most likely related
  to the precession of the radio jet of \lsi{}, see the agreement with
  the precessional period from VLBA astrometry, of 27--28~days
  \citep{Massi2012}. The hypothesis that a precessing jet, with an
  approaching jet with large excursions in its position angle, gives
  rise to appreciable Doppler boosting effects (and therefore to
  changes of flux density that can be detected by timing analysis) is
  supported by the morphology of images reported by
  \citet{Massi2004a}, \citet{Dhawan2006}, and \citet{Massi2012} that
  showed extended radio structures changing from two-sided to
  one-sided morphologies at different position angles. A physical
  model for \lsi{} of synchrotron emission from a precessing ($P_2$)
  jet, periodically ($P_1$) refilled with relativistic particles, has
  shown that the maximum of the long-term modulation occurs when $P_1$
  and $P_2$ are synchronized, that is the jet electron density is at
  about its maximum and the approaching jet forms the smallest
  possible angle with the line of sight. This coincidence of the
  highest number of emitting particles and the strongest Doppler
  boosting of their emission occurs with the frequency of
  $\nu_1-\nu_2$ and creates the long-term modulation observed in
  \lsi{} \citep{Massi2014}.
}
\item{
  The folded curves of $\gamma$-ray data show that the peak at
  apoastron seems to be affected by the same orbital shift that
  affects the radio outburst. When the radio outburst occurs at
  orbital phase $\Phi \sim 0.65$, the first main GeV peak at
  periastron shows a bump at orbital phase $\Phi \sim 0.65$. When the
  radio  outburst is shifted towards orbital phase $\Phi \approx 0.9$,
  the apoastron peak appears at the same orbital phases and is
  detached from the first main GeV peak.
}
\end{enumerate}

We conclude that there exists a GeV peak at apoastron with the same
timing characteristics and orbital shift as the radio emission. 
Because of the orbital shift, 
at some $\Theta$-intervals
this GeV peak
is detached enough from the periastron $\gamma$-ray peak 
to become
discernable as a second peak, 
 and then it influences the timing
analysis. Future analyses should investigate how the radio outburst
and second GeV peak are connected;
constraints to these future analyses should result from a detailed
investigation of the evolution of the spectrum with orbital phase to
see the increase of some spectral features at various orbital phases.

%
%
%
%

\begin{acknowledgements}
  We thank Bindu~Rani for reading the manuscript and useful comments,
  Robin Corbet for answering questions concerning the reduction of
  \textit{Fermi}-LAT data and Chris Torrence for answering questions
  concerning the wavelet analysis. We thank Walter~Alef and
  Alessandra~Bertarini for their support with the computing
  power. Finally, we thank the anonymous referee for the careful
  reading of our manuscript and the useful comments made for its
  improvement. The wavelet software was provided by C.~Torrence and
  G.~Compo, and is available at URL:
  http://atoc.colorado.edu/research/wavelets/. The Green Bank
  Interferometer was operated by the National Radio Astronomy
  Observatory for the U.S.\ Naval Observatory and the Naval Research
  laboratory during the timeperiod of these observations. This work
  has made use of public \textit{Fermi} data obtained from the High
  Energy Astrophysics Science Archive Research Center (HEASARC),
  provided by NASA Goddard Space Flight Center.
\end{acknowledgements}



\end{document}